# Incorporating Trip Chaining within Online Demand Estimation

Guido Cantelmo[a*], Moeid Qurashi[a], A. Arun Prakash[c], Constantinos Antoniou[a], Francesco Viti[b]

[a]*Technical University of Munich, 80333 Munich, Germany*
[b]*University of Luxembourg, L-4365 Esch-sur-Alzette, Luxembourg*
[c]*Massachusetts Institute of Technology, Cambridge, MA 02139-4307, USA*

**Abstract**

Time-dependent Origin-Destination (OD) demand flows are fundamental inputs for Dynamic Traffic Assignment (DTA) systems and real-time traffic management. This work introduces a novel state-space framework to estimate these demand flows in an online context. Specifically, we propose to explicitly include trip-chaining behavior within the state-space formulation, which is solved using the well-established Kalman Filtering technique. While existing works already consider structural information and recursive behavior within the online demand estimation problem, this information has been always considered at the OD level. In this study, we introduce this structural information by explicitly representing trip-chaining within the estimation framework. The advantage is twofold. First, all trips belonging to the same tour can be jointly calibrated. Second, given the estimation during a certain time interval, a prediction of the structural deviation over the whole day can be obtained without the need to run additional simulations. The effectiveness of the proposed methodology is demonstrated first on a toy network and then on a large real-world network. Results show that the model improves the prediction performance with respect to a conventional Kalman Filtering approach. We also show that, on the basis of the estimation of the morning commute, the model can be used to predict the evening commute without need of running additional simulations.

## 1. Introduction

The ever-growing demand pressure on existing transport facilities generates significant societal, environmental and economic losses for our society. To tackle this issue, practitioners and researchers rely on traffic models to evaluate and implement efficient measures, which range from off-line planning to real-time traffic management solutions. However, the efficiency of these tools depends on their ability to correctly predict traffic conditions. In this context, one of the most important inputs is the mobility demand, which is usually discretized in the form of Origin-Destination (OD) demand matrices. In transport engineering, estimating these demand matrices using traffic counts is a classical and widely adopted procedure (Cascetta, Inaudi, and Marquis 1993). This problem, referred to as Dynamic OD demand Estimation (DODE) in literature, seeks for the best possible approximation of OD flows that minimises the error between simulated and available traffic data. Traditionally, this procedure can be applied off-line (medium-long term planning) (Cascetta, Inaudi, and Marquis 1993) or on-line (real-time traffic prediction) (Antoniou, Ben-Akiva, and Koutsopoulos 2007).

The DODE is prone to exhibit underdeterminedness because of the high number of unknown variables in comparison to the number of observations (Marzano, Papola, and Simonelli 2009; Prakash et al. 2018). As a resolution, many authors assume that good starting values of the demand (seed values) are available and are directly included in the estimation process (Ashok and Ben-Akiva 2000; Cipriani et al. 2014). However, for complex, congested networks this condition becomes a restriction, meaning that the DODE procedure results in a suboptimal solution (Marzano, Papola, and Simonelli 2009; Zhang et al. 2018).

Over the last decade, substantial research has been carried out to tackle this issue. However, recent works in both on-line and off-line demand estimation mostly focus on adopting dimensionality reduction techniques, such as Principal Component Analysis (PCA) (Djukic, van Lint, and Hoogendoorn 2012; Prakash et al. 2018). Although these techniques can be efficient, they rely on statistical rules that introduce intrinsic errors related to the variables that have not been properly calibrated (Cascetta et al. 2013).

An alternative approach is to leverage behavioural models to reduce the number of variables without introducing intrinsic errors (Cantelmo et al. 2018a). Conventional DODE models treat OD pairs as independent and uncorrelated variables (Cascetta, Inaudi, and Marquis 1993); however, mobility demand derives from the way individuals plan their activities (Djukic, van Lint, and Hoogendoorn 2012). Consequently, the intuition is to include structural information about the demand and its composition within the DODE to reduce the underdeterminedness of the problem. While this approach has been applied to the off-line DODE (Lindveld 2003; Flötteröd, Chen, and Nagel 2012; Cantelmo et al. 2018a), —to the best of the authors' knowledge— there is no existing framework for the on-line case that accounts for structural information about trip chain within the estimation framework.

This paper contributes to the state of the art on online demand estimation by introducing trip chaining behaviour within the on-line DODE. Specifically, we combine the Utility-Based formulation proposed in Cantelmo et al. (2018) with a state-space framework for real-time OD estimation. By explicitly considering trip chaining behaviour, we show that the demand estimated in earlier time periods, e.g. in the morning rush hour, can be used to improve the prediction quality during later time periods, e.g. the evening commute, and hence ensure consistency over a 24-hours period. Specifically, we include trip chaining information within the well-established Kalman Filtering (KF) online demand estimation approach, to achieve a more reliable estimation while keeping the computational time suitable for real-time applications. Hence, the proposed formulation brings two main contributions with respect to the state of the art: first, we create a model that is more robust and consistent over the day, as the activity-based demand structure is included within the estimation framework. Secondly, the framework can be applied for studying the evolution of the demand, hence capturing activities' evolution over a day.

## 2. Literature Review

### 2.1. Dynamic OD Estimation approaches

Estimation and prediction of dynamic demand flows is a classical and widely common procedure in transportation engineering, which is conventionally studied in two contexts: offline (transport planning) and online (real-time management) applications. While for a detailed overview of existing frameworks we refer the interested reader to other works (Antoniou et al. 2016; Carrese et al. 2017), in this section we provide a short overview on the most relevant related works.

The offline approach is usually formulated as an optimization problem and solved through a Generalized Least Squared (GLS) approach (Cascetta, Inaudi, and Marquis 1993; Zhang, Nie, and Qian 2008). However, this formulation provides poor results if historical information about the OD flows is not included within the GLS objective function (Zhang, Nie, and Qian 2008; Marzano, Papola, and Simonelli 2009). Even in this case, the DODE remains a non-linear problem and it is likely to converge to a sub-optimal solution. One option to overcome this limitation is to reduce the number of decision variables. In this sense, many alternative approaches have been proposed. Djukic et al. (2012) proposed to use Principal Component Analysis (PCA) to identify the most important elements of the problem and remove those that create noise in the objective function. Cascetta et al. (2013) propose to reduce the number of decision variables by working on the generated demand – i.e. the total demand flow leaving each traffic zone within a certain time interval - instead of the individuals OD pairs. Lindveld (2003) proposes a similar approach, where a departure time choice model is adopted to distribute demand flows over time. Similarly, Cantelmo et al. (2015) propose to use a sequential approach that first estimates the generated demand for each traffic zone and in a second phase adjusts the disaggregate OD flows in order to achieve a more reliable estimation of the traffic data. Another possibility is to leverage traffic data to learn the structure of the demand (Antoniou, Ben-Akiva, and Koutsopoulos 2004; Zhao et al. 2010; Barceló Bugeda et al. 2012; Nigro, Cipriani, and Giudice 2018). As including this information within the conventional GLS framework can be challenging, Balakrishna et al. (2007) proposed to use the more general Simultaneous Perturbation Stochastic Perturbation (SPSA), which employs a Dynamic Traffic Assignment (DTA) model as a black box in order to generate traffic data, instead of using complex mathematical formulations. Additionally, the authors pointed out how supply and demand parameters should be jointly calibrated in order to escape from local minima. Similarly, Cantelmo et al. (2018) proposed an SPSA-based approach to estimate the parameters of a departure time choice model instead of the disaggregate OD flows. The advantage in this case is that the model can explicitly account for different activity patterns while reducing the complexity of the OD estimation problem. However, computational time remains an issue, as the model needs to consider a 24-hours scenario in order to fully exploit activity information to estimate OD trips.

The most established method for online DODE is to reformulate the problem as a state-space model and, successively, to adopt a Kalman Filtering (KF) approach (Okutani and Stephanedes 1984). In order to include the structure of the demand within the estimation framework, Ashok and Ben-Akiva (2000) formulated the KF in term of deviations between actual and historical demand flows. To achieve a more reliable representation of the demand, the authors further extended the KF model in order to account for the stochasticity in the demand flows (Ashok and Ben-Akiva 2002). However, while the KF provides excellent estimations for linear systems, DTA is a complex tool characterized by highly non-linear and correlated functions. To overcome this limitation, different solutions have been proposed, including the Extended KF (Antoniou, Ben-Akiva, and Koutsopoulos 2007) and linear approximations of the demand flows (Zhou and Mahmassani 2007). A major issue for the real-time demand estimation is the inability of the KF to handle large numbers of variables (Bierlaire and Crittin 2004). Again, two main paths have been explored in past literature. On one hand, the inclusion of additional information within the DODE, (Antoniou, Ben-Akiva, and Koutsopoulos 2004; Zhang, Nie, and Qian 2008; Barcelo et al. 2013; Carrese et al. 2017). On the other hand, reduction of the problem's dimensionality. Prakash et al. (2018) combined PCA and the KF in order to improve the scalability of the model with respect to the conventional approach. (Marzano et al. 2015, 2018) combined the Quasi-dynamic approach proposed in Cascetta et al. (2013) with a Kalman Filter, showing that structural information can also be combined with the KF model. However, there is no research so far attempting to include information at tour level within the online DODE or within the Kalman Filtering framework. While existing works show that this can bring substantial advantages in the offline case (Lindveld 2003; Flötteröd, Bierlaire, and Nagel 2011; Cantelmo et al. 2018a), we argue in this paper that considering trip chaining in an online framework is a fundamental step in order to bring consistency in the demand and enhance the prediction capabilities of the state space model.

*2.2. Trip-Chaining and aggregate demand models*

Conventional DODE approaches do not consider any correlation between demand flows, either in time or space. However, OD flows are an aggregated representation of individuals' activity-travel chains.

Activity chains are traditionally modelled through Activity-Based demand Models (ABM), which use disaggregate information (such as travel diaries) to represent user behavior at an individual or household level (Bowman and Ben-Akiva 2001; Viegas de Lima et al. 2018). Among the different ABM applications, a classical approach is to use ABM models to calibrate a departure time choice (activity scheduling) model that reflects user preferences in terms of activity scheduling and behavior (Ettema and Timmermans 2003). The idea is that, starting from a sample of the population, mathematical relationships that correlate the departure time with the utility of performing a certain activity at a certain time can be derived. Following this line, an intensive research effort has been carried on to exploit these ABM departure time choice models within both disaggregate (Feil, Balmer, and Axhausen 2009) and aggregate (Adnan 2010) DTA models. However, while in both cases the advantage of linking activity chaining and DTA provides useful insights about user behavior for planning purposes (such as dynamic tolling), the calibration of these models is challenging especially in large scale applications (Flötteröd, Bierlaire, and Nagel 2011). To be more precise, the main problem is not within the ABM methodology itself, which in principle can handle large networks, but within the data that are usually limited in size and can hardly cover large metropolitan areas. For this reason, traffic data, and specifically traffic counts, can also be used to calibrate and improve these ABM models on large networks (Flötteröd, Bierlaire, and Nagel 2011). However, this is still usually done offline to calibrate behavioral models, which are then mostly employed for planning purposes. By contrast, our proposed model uses the aggregate approach for real-time traffic prediction. Hence, the proposed methodology builds upon flow-based ABM departure time choice models — such as the one proposed in Adnan (2010)— which use non-linear utility functions to model purpose dependent OD flows. Inspired by this modeling approach, some of the authors of this paper have recently proposed a DTA model and an offline demand estimation framework that explicitly accounts for departure time choice and activity scheduling and duration. We refer the interested reader to Cantelmo and Viti (2018); Cantelmo et al. (2018a) for more detail on adapting ABM departure time choice models to model OD flows.

## 3. Methodology

Following the works of Zhou and Mahmassani (2007) and Antoniou et al. (1997) the structure of the demand is considered in this research as the convolution of three components: *Regular Demand Patterns*, *Structural Deviations* and *Random Fluctuations*. Regular Demand Patterns represent the typical underlying demand profile, Structural Deviations take into account phenomena such as weather conditions, which the analyst can model or account for explicitly through e.g. additional information. Random fluctuations consider instead those deviations that cannot be explained and are caused by the stochasticity of the estimated variables and the collected data. In the proposed methodology, we include a parametric function within the KF in order to account for trip chaining within the online framework. This approach will be used to update the structural component of the demand, while the standard KF proposed in Ashok and Ben-Akiva (2000) is adopted to capture the regular and random components.

### 3.1. The historical demand matrix: Purpose-dependent OD flows

Before introducing the online calibration problem and its formulation, in this section we briefly discuss the role of the historical matrix, the *structural information*, and the *regular demand patterns*. The main idea is that homogeneous classes of users (classified e.g. by common trip purpose) might be representable through a small set of carefully selected parameters. Simply stated, it should be possible to model users travelling from a certain origin to a certain destination for a certain purpose through a limited number of parameters, such as the preferred expected departure or arrival times and their corresponding variance.

Fig. 1 provides an illustrative example of aggregated daily demand flows over a whole day and for an OD pair as simple convolution of Gaussian functions, each representing a purpose-specific demand pattern. Therefore, each function is characterized by a mean departure or arrival time μ and a standard deviation σ. This assumption has the notable advantage that OD flows are modelled as continuous in time and differentiable, and that their parameters are easily interpretable.

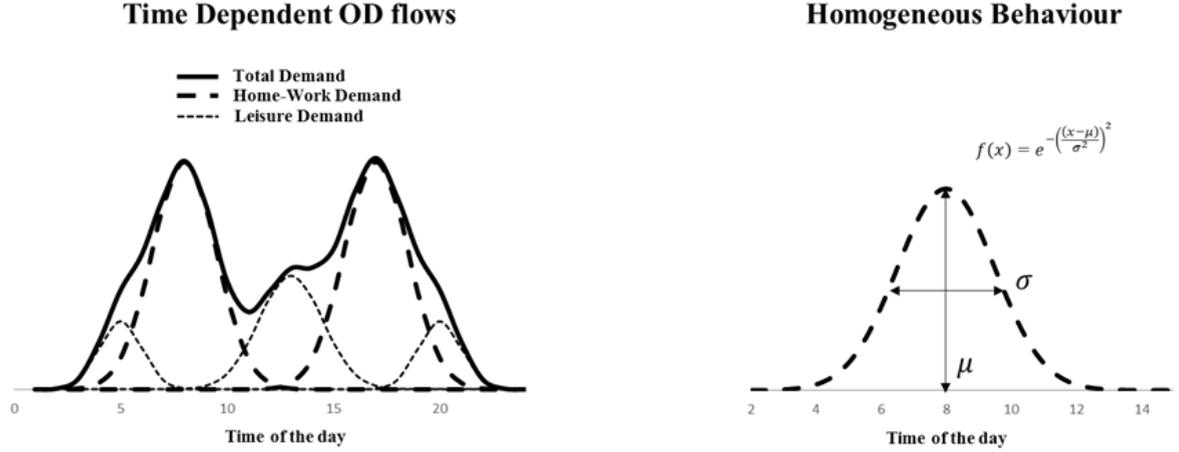

Fig. 1 Mobility Demand as a convolution of different activities

Being aware that simple gaussian functions do not represent the departure time in a realistic way, each trip can be modelled through (nonlinear-in-parameters) utility functions. While identifying and characterizing these functions is beyond the scope of this paper, having this information is a requirement in order to include trip chaining within the online prediction model. Looking at the ABM literature, the first key concern is determining how a theory that has been developed to model individual user behavior can be combined with a framework – the OD estimation – that relies on a highly aggregate representation of the demand. As mentioned in the previous sub-section, there are some ABM applications, mostly developed in the domain of transport economics, that focus on combining discrete choice theory and utility maximization theory to model the departure time choice at an aggregate level. Let us consider an analysis period $T$ which is divided in $n_h$ time intervals, a network composed of $n_{od}$ OD pairs, and a number of trip purposes $n_p$. In principle, given a set of utility functions, their parameters $\boldsymbol{\theta}$, and the vector $\mathbf{tt}$, with dimension $n_{od} \times n_h$, describing the travel time for each OD and each time interval, flows of vehicles are loaded onto the network using fixed-point solution algorithms (such as Method of Successive Averages or the Frank-Wolfe algorithm) until convergence is reached (Adnan 2010; Cascetta 2009). Vector $\boldsymbol{\theta}$ has dimension $n_p \times n_\theta$, with $n_\theta$ the number of parameters of the utility function. Convergence follows usually the Wardrop 1st equilibrium principle, i.e. when the utility of all used time intervals is equal or less than those that would be experienced by a single user on any different time interval, convergence is achieved (Cantelmo and Viti 2018). As the model adopts a different utility function for each different trip purpose $p$, we can have that the output of such a framework is a probability function $P_h^p(\boldsymbol{\theta}, \mathbf{tt})$ of choosing time interval $h$, which depends on the trip purpose $p$, the parameters of the function describing the utility of performing that activity $\boldsymbol{\theta}$, and the $\mathbf{tt}$ travel times at time interval $h$.

Different approaches have been proposed to obtain functions —such as those presented in Fig.1 for general networks— depending on the available input data. If no time-dependent historical OD matrix is available, the Utility-Based OD Estimation (UB-DODE) framework proposed in Cantelmo et al. (2018) can jointly estimate offline both activity-based OD flows and parameters $\boldsymbol{\theta}$ of the utility function. Alternatively, if an historical (dynamic) OD matrix is already available, a Markov Chain Monte Carlo model can be used to approximate $P_h^p(\boldsymbol{\theta}, \mathbf{tt})$ for each traffic zone, as proposed in Scheffer et al. (2017). Another option is to use activity based demand modelling (Bowman and Ben-Akiva 2001) and/or to generate a synthetic population for the study area (Barthelemy and Toint 2012). However, in this case, a sample of the population or additional socio-demographic data are required (e.g. via a travel diary data collection). This type of solution already exists in some countries, such as Switzerland (Balmer et al. 2008).

Irrespective of the chosen approach, the output is, for each OD pair $i$ and trip purpose $p$, the estimated number of users travelling from a certain origin to a certain destination $N_p^i$ and the probability of departing during a specific time interval $h$. Once the probability function has been properly specified, the expected demand value can be expressed as:

$$X_h^i = \sum_{p \in P} N_p^i P_h^{i,p}(\boldsymbol{\theta}, \mathbf{tt}) \tag{1}$$

where $X_h^i$ is the OD demand flow for OD pair $i$ at time interval $h$, $P_h^{i,p}$ is the probability of departing during time interval $h$ for OD pair $i$ and for performing a trip purpose $p$; $\boldsymbol{\theta}$ represents the vector of parameters for the utility function describing that specific activity (related to departure time to perform an activity and duration of the said activity) and $\mathbf{tt}$ represents the vector of OD trip times, which also influences the travel behavior.

*3.2. Trip chaining and demand legs*

Online estimation of all parameters composing Equation (1) might be an option, however, some of them are redundant when the goal is to predict the OD flows over a short period of time. For instance, $\boldsymbol{\theta}$ is mostly responsible for describing activity preferences and scheduling, which we can assume to be constant over a given day. Instead, the parameter $N_p^i$ plays a fundamental role, as the demand for a specific activity can drastically change from one day to the next (e.g. a planned football match or a bank holiday). Moreover, demand for different activities is interconnected and can be used for prediction purposes (such as morning commute and evening commute) or to forecast the impact of congestion pricing schemes (Zockaie et al. 2015).

In order to exploit this concept within the online OD estimation, we divide the demand for the study period, for instance 24 hours, in several matrices called *legs*, where each cell of each matrix represents the demand $N_p^i$ for a certain activity $p$ and a certain OD pair $i$, while $\mathbf{N_p}$ represents a $n_{od} \times n_P$ matrix whose columns are the OD flows for the activity-specific demand pattern $p$. To provide an example, let assume that the home-work daily pattern can be represented through two demand matrices (or legs), one representing the demand for the morning commute and the other representing the demand during the evening commute.

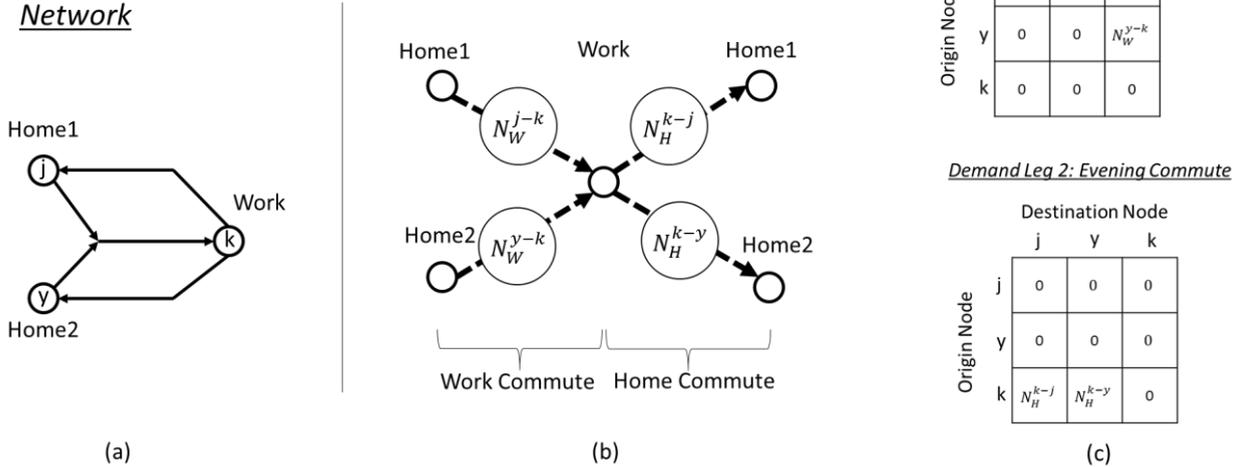

Fig. 2: (a) Synthetic Network; (b) Graphical representation of the relationship among legs; (c) Matrices representing the two demand legs for the activity pattern (Home-Work-Home);

From this representation, functional relations can be calculated in order to understand how OD flows are structurally related over time and space. The concept is demonstrated using the illustrative example reported in Fig. 2, where two zones (*j-y*) are assumed to be points of origin for the working commute towards a single zone of destination *k* (Fig.

2a). $N_W^{j-k}$ and $N_W^{y-k}$ represent the flows moving from the 'home' zones to the 'work' zone, while $N_H^{k-j}$ and $N_H^{k-y}$ are those from the working zone observed to return later in time to the residential zones. This implies that $(N_W^{j-k} + N_W^{y-k} = N_H^{k-j} + N_H^{k-y})$, considering conservation of vehicles. This simple example illustrates how variables $N_p^i$ are structurally related.

We can formulate these structural relations mathematically as follows. Let us define $A_W = (N_W^{j-k} + N_W^{y-k})$ the total demand to node $k$ during the morning commute and $G_H = (N_H^{k-j} + N_H^{k-y})$ the total demand from node $k$ during the evening commute. By initially assuming strict conservation of demand flows, we can write $A_W = G_H$ – i.e. we expect that all travelers leaving home in the morning will return home by the end of the day – which leads to the following system of equations:

$$\begin{cases} N_W^{j-k} + N_W^{y-k} = A_W \\ N_H^{k-j} + N_H^{k-y} = G_H \\ G_H = A_W \\ N_H^{k-j} = A_W \left( \dfrac{N_H^{k-j}}{G_H} \right) \\ N_H^{k-y} = A_W \left( \dfrac{N_H^{k-y}}{G_H} \right) \end{cases} \quad (2)$$

If real-time demand estimation is performed on this system, after some time a new value of the OD flows for the morning commute, $N^*{}_W^{j-k}$ and $N^*{}_W^{y-k}$, will be available. The total flow travelling to node $k$ in the morning commute can be calculated as $A_W^* = \left( N^*{}_W^{j-k} + N^*{}_W^{y-k} \right)$. After the new estimation of $A_W^*$ has been obtained, the system of Equations (2) will provide an estimation of the OD flows in the afternoon, which can be calculated as:

$$\begin{cases} N^*{}_H^{k-j} = A_W^* \left( \dfrac{N_H^{k-j}}{G_H} \right) \\ N^*{}_H^{k-y} = A_W^* \left( \dfrac{N_H^{k-y}}{G_H} \right) \end{cases} \quad (3)$$

meaning that OD flows travelling in the morning are used to predict the traffic in the evening.

We now further generalize this concept for more realistic systems. Let us define $G_p^n$ as the total demand leaving traffic zone $n$ for purpose $p$, where $D_{n,p}^i = \dfrac{N_p^i}{G_p^n}$ is the fraction of this demand that travels between OD pair $i$. By taking into account that $n$ is the origin node for OD pair $i$, we define $A_{p^k<p}^n$ the overall demand going to $n$ explained through a previous demand leg $p^k$. As the demand for a certain activity can be function of different demand segments (i.e. several activities could have the same common destination), the set of Equations (2) can be generalized as:

$$N_p^i = \left( \sum_{k \in A} A_k^n \right) D_{n,p}^i \quad (4)$$

Where $A$ is the set of previous demand legs contributing to the demand for the current leg – or activity – $p$. If for instance both workers and students are travelling to node $n$ in the morning, we will have $\sum_{k \in A} A_k^n = A_W^n + A_{School}^n$. It is clear then that the number of activities have to be consistent with equation (4) – i.e. multiple demand matrices for the same activity purpose can be generated in order to ensure consistency for different activity patterns.

*3.3. State-space formulation*

In this section we briefly introduce the concept of state-space modeling for online demand estimation. We refer the interested reader to (Antoniou, Ben-Akiva, and Koutsopoulos 2007) for more details.

The online DODE problem formulated as a state-space model is composed of three main elements:
1. The state vector, – i.e. the set of variables that is sufficient to uniquely describe the evolution of the system;
2. The transition equation, which describes the evolution of the system over time;
3. The measurement equation, which maps the available traffic data to the state vector.

Following Ashok and Ben-Akiva (2000), we formulate the state-space model in terms of traffic state deviation. Again, the reason is that by using deviations instead of demand flows, the model takes into account the historical structure of the demand matrix during the estimation. Let $\mathbf{x}_h$ represents the vector containing the demand parameters of the DTA model at time interval $h$. In this case, the demand parameters to be calibrated are the OD flows. Let us define $\mathbf{x}_h^{hist}$ the matrix $n_{od} \times 1$ with the historical value of the OD flows, before the calibration. The state vector is then indicated as $\Delta \mathbf{x}_h = \mathbf{x}_h - \mathbf{x}_h^{hist}$. As proposed in Ashok and Ben-Akiva (2000), the transition equation is approximated through the autoregressive process detailed in Equation (5):

$$\Delta \mathbf{x}_h = \sum_{k=h-T}^{h-1} \mathbf{F}_k^h \Delta \mathbf{x}_i + \mathbf{u}_h \quad (5)$$

Where $T$ represents the number of previous time intervals that influence the demand at time $h$, $\mathbf{F}_k^h$ is the $n_{od} \times n_{od}$ vector of autoregressive factors relating the parameters estimated at time interval $k$ to those estimated during the current time interval $h$, and $\mathbf{u}_h$ is the $n_{od} \times 1$ vector of random errors in the transition equations. Equation (5) represents thus the "expected" evolution of the system based on historical information.

The measurement equation measures instead the influence of the additional available information – i.e. the traffic counts – on the system. Instead of traffic counts, and to be consistent with the state vector, the measurement equation is also formulated in term of deviations between available and historical traffic counts as in Equation (6):

$$\Delta \mathbf{y}_h = \sum_{k=1-T}^{h} \mathbf{H}_k^h \Delta \mathbf{x}_k + \mathbf{v}_h \quad (6)$$

Where $\Delta \mathbf{y}_h = \mathbf{y}_h^{sim} - \mathbf{y}_h^{hist}$ is the $n_l \times 1$ vector of the deviations between historical traffic counts $\mathbf{y}_h^{hist}$ and simulated ones $\mathbf{y}_h^{sim}$ at time interval $h$, with $n_l$ number of links. $\mathbf{H}_k^h$ is the $n_l \times n_{od}$ assignment matrix, which describes the relationship between the state vector $\Delta \mathbf{x}_h$ and the measurement vector $\Delta \mathbf{y}_h$.

The Kalman Filtering approach assumes that at each step we have two Gaussian distributions: the predicted state, with mean equal to Eq. (5) and variance equal to Eq. (11), and the observed data, with mean values equal to Eq. (6) and variance Eq. (7).

$$\mathbf{\Sigma}_h = \sum_{k=h-T}^{h} \mathbf{H}_k^h \mathbf{\Sigma}_k \mathbf{H}_k^{h^T} + \mathbf{R}_h \quad (7)$$

where $\mathbf{P}_h$ represents the $n_{od} \times n_{od}$ covariance matrix of the estimates of the state vector $\Delta \mathbf{x}_h$, $\mathbf{\Sigma}_h$ is the covariance matrix of the measurement equations, and $\mathbf{Q}_h/\mathbf{R}_h$ are vector of random errors. Then, the Kalman Filter determines the most likely value for the state vector by considering the combination of these two distributions. The main steps of the algorithm are described below:

| Algorithm 1. Kalman Filter |
|---|
| Initialization at time interval 0 |
| $$\Delta \mathbf{x}_{0|0} = \Delta \mathbf{x}_0 \quad (8)$$ $$\mathbf{P}_{0|0} = \mathbf{P}_0 \quad (9)$$ |
| Time update for the current time interval $h$ |
| $$\Delta \mathbf{x}_{h|h-1} = \sum_{k=h-T}^{h-1} \mathbf{F}_k^h \Delta \mathbf{x}_i \quad (10)$$ $$\mathbf{P}_{h|h-1} = \sum_{k=h-T}^{h-1} \mathbf{F}_k^h \mathbf{P}_{k|k} \mathbf{F}_k^{h^T} + \mathbf{Q}_h \quad (11)$$ |
| Measurement update for the current time interval $h$ |
| $$\mathbf{K}_h = \mathbf{P}_{h|h-1} \mathbf{H}_h^T \left( \mathbf{H}_h \mathbf{P}_{h|h-1} \mathbf{H}_h^T + \mathbf{Q}_h \right)^{-1} \quad (12)$$ $$\Delta \mathbf{x}_{h|h} = \Delta \mathbf{x}_{h|h-1} + \mathbf{K}_h \left( \Delta \mathbf{y}_h - \mathbf{H}_h \Delta \mathbf{x}_{h|h-1} \right) \quad (13)$$ $$\mathbf{P}_{h|h} = \mathbf{P}_{h|h-1} - \mathbf{K}_h \mathbf{H}_h \mathbf{P}_{h|h-1} \quad (14)$$ |

In the algorithm, $\Delta \mathbf{x}_{h|h-1}$ and $\mathbf{P}_{h|h-1}$ are the predictions based on the transition equation alone, and $\mathbf{K}_h$ represents the so-called "Kalman Gain". By combining the Kalman Gain with the prediction from the transition equation, we can estimate $\Delta \mathbf{x}_{h|h}$ and $\mathbf{P}_{h|h}$, which represent the most likely values of the state vector according to both the measurement and transition equations.

*3.4. Introducing trip chaining relations within the State-Space model*

The standard KF algorithm formulated as in (Ashok and Ben-Akiva 2000) and presented in Section 3.3 is a powerful tool but limited to linear systems. However, non-linear functions are more realistic in capturing the evolution of the demand over time, as the travel demand is characterized by different interacting mobility patterns. If a realistic profile of the demand is available, such as the one discussed in Section 3.1, and considering that Equation (5) includes the deviations between estimated and historical demand, the State-Space model formulated in the previous section can provide reliable predictions on the demand at time interval $h$. However, this estimation is based only on local changes in the demand, which can stem from random fluctuations, while the main purpose of this work is to consider trip chaining and interdependencies between demand flows belonging to different demand legs in order to better capture the dynamics of the structural component. To achieve this goal, this subsection introduces an extension of the Kalman Filter – named Parametric Kalman Filter (PKF) – in order to explicitly consider these dynamics.

Define $\mathbf{N}_p^{est}$ the vector $n_{od} \times 1$ of estimated demand flows for demand leg (or trip purpose) $p$, while $\mathbf{N}_p^{hist}$ the vector of historical demand values (i.e. the static OD matrix including purpose dependent demand flows for each OD pair). First, we formulate the state vector as the difference between estimated and historical OD flows for all demand legs $\Delta \mathbf{N}_p = \mathbf{N}_p^{est} - \mathbf{N}_p^{hist}$. The transition equation can be thus formulated as:

$$\Delta \mathbf{N}_p = \sum_{k \in A} \mathbf{D}_p \Delta \mathbf{N}_k + \mathbf{e}_p \quad (15)$$

where the $\mathbf{D}_p$ is the $n_{od} \times 1$ vector including the demand fractions for the demand leg $p$ calculated as in Equation (4), $A$ is the set of OD flows of all previous demand legs influencing the current leg. Finally, $\mathbf{e}_p$ is a vector of error terms.

Equation (15) aims to properly capture structural deviations and relations among demand flows. For instance, if an excessive demand is observed in the morning, Equation (15) will take this into account for predicting the afternoon demand. It should be stressed out that the proposed methodology is not limited to the simple home-work commute but to any trip chain, as long as $\mathbf{D}_p$ has been properly calibrated before running the online prediction model. However, this model will also propagate eventual errors in one demand leg to the others since the estimated time-dependent OD flows are functionally related. To correct this issue, also in this case we define a measurement equation in order to achieve more reliable predictions. Specifically, the measurement equation in this case will be the same as the one proposed in (6), with the difference that we adopt the cumulative of the link flows over a reference period instead of the link flows during a certain time interval $h$. If the cumulative of the estimated and historical flows is the same, this approach will avoid transferring excessive amounts of demand information from one demand leg to the other:

$$\Delta \mathbf{Y}_h = \sum_{i \in P} \sum_{k=1-T}^{h} \mathbf{A}_{k,i}^{h} \Delta \mathbf{N}_i + \mathbf{v}_h = \sum_{i \in P} \sum_{k=1-T}^{h} \mathbf{H}_k^h \mathbf{T}_k \Delta \mathbf{N}_i + \mathbf{v}_h \qquad (16)$$

where $\Delta \mathbf{Y}_h$ is the $n_l \times 1$ matrix with the difference between the cumulative of the actual and historical link flows at time interval $h$, which is the time interval in which we update the demand value for the leg $p$. $\mathbf{A}_{k,i}^{h}$ maps the state vector $\Delta \mathbf{N}_p$ to the link flows and it can be obtained by multiplying the assignment matrix and the probability derived from the departure time choice model $P_h^{i,p}$. In Equation (16), $\mathbf{T}_k$ is a $n_{od} \times n_{od}$ matrix with all zero elements, except along the main diagonal, which contains the probability $P_h^{i,p}$ for OD $i$ to depart during time interval $h$.

Let us now define $\mathbf{P}_p^{Leg}$ as the covariance matrix of the estimates of $\Delta \mathbf{N}_p$ and $\mathbf{Q}_p^{Leg}$ the vector of error terms. The full system of equations defining the Parametric Kalman Filter (PKF) yields thus to the following Algorithm 2:

| Algorithm 2. Parametric Kalman Filter |
|---|
| Initialization at time interval 0 |
| $\Delta \mathbf{N}_{0\|0} = \Delta \mathbf{N}_0 \qquad (17)$ |
| $\mathbf{P}_{0\|0}^{Leg} = \mathbf{P}_0^{Leg} \qquad (18)$ |
| Time update for the current demand leg $p$ |
| $\Delta \mathbf{N}_{p\|p-1} = \sum_{k \in A} \mathbf{D}_p \Delta \mathbf{N}_{k\|k} \qquad (19)$ |
| $\mathbf{P}_{p\|p-1}^{Leg} = \sum_{k=h-T}^{h-1} \mathbf{D}_k^h \mathbf{P}_{k\|k}^{Leg} \mathbf{D}_k^{h^T} + \mathbf{Q}_p^{Leg} \qquad (20)$ |
| Measurement update for the current demand leg |
| $\mathbf{K}_p = \mathbf{P}_{p\|p-1}^{Leg} \mathbf{A}_p^{h^T} \left( \mathbf{A}_p^h \mathbf{\Sigma}_{p\|p-1}^{Leg} \mathbf{A}_p^{h^T} + \mathbf{Q}_p^{Leg} \right)^{-1} \qquad (21)$ |
| $\Delta \mathbf{N}_{p\|p} = \Delta \mathbf{N}_{p\|p-1} + \mathbf{K}_p \left( \Delta \mathbf{Y}_h - \sum_{i \in P} \sum_{k=1-T}^{h} \mathbf{A}_{k,i}^h \Delta \mathbf{N}_i \right) \qquad (22)$ |
| $\mathbf{P}_{p\|p}^{Leg} = \mathbf{P}_{p\|p-1}^{Leg} - \mathbf{K}_p \mathbf{A}_p^h \mathbf{P}_{p\|p-1}^{Leg} \qquad (23)$ |

As for the classical KF, the model first estimates the more likely demand flows based only on the transition equation – i.e. the information about trip chain derived from the transition probabilities discussed in sub-section 3.2 – and then exploits the Kalman Gain in order to improve the estimation through the traffic counts.

The main novelty lies in formulating the state-space model as a function of the activities by introducing (i) a new state vector $\Delta \mathbf{N}_p$, (ii) a new transition equation (15), and (iii) a new measurement equation (16). Another major difference between the PKF and KF is in the time updating phase. While the KF updates the demand from one time interval to the next, the PKF updates the demand from one leg – or activity – to the other. In other words, the PKF models the transition from one OD to the other taking into account trip chaining, while the KF models the evolution of the demand over time. For example, let us say that a traffic center wants to predict demand during the afternoon. If a significant change is observed with respect to the historical demand, based on the traffic counts and the demand estimated during the morning, the PKF can be used to predict the demand flows for all time-intervals during the day, without needing additional data or additional simulation. The traffic center can thus immediately use the PKF to predict atypical user behavior rather than waiting for real-time traffic data in order to compute the prediction through the standard KF approach.

*3.5. Prediction of the demand flows and practical considerations*

The conventional Kalman Filter (KF) and the novel Parametric Kalman Filter (PKF), introduced in Sections 3.4 and 3.5, respectively, capture two different errors. The KF directly corrects demand flows based on the traffic counts at the current time interval. The PKF leverages estimates from all previous time intervals to correct the purpose-specific demand flows of the next time intervals. Thus, in a real time setting, both ingredients have to be properly included during the prediction phase. Hence, in order to faithfully estimate the demand, the following equation should be adopted:

$$X_h^i = X_h^{i,hist} + \sigma_{KF}^i + \sigma_{PKF}^i \tag{24}$$

Where $X_h^i$ is the estimated demand for OD pair $i$ at time interval $h$, $X_h^{i,hist}$ is its historical value, $\sigma_{KF}^i$ is the random deviation estimated through the KF (Algorithm 1), and $\sigma_{PKF}^i$ accounts for the structural and regular components related to the PKF (Algorithm 2).

Equation (24) can be written directly as a function of the outputs from Algorithms 1&2 as:

$$X_h^i = X_h^{i,hist} + \Delta X_h^i + \left( \sum_{p \in P} \Delta N_p^i P_h^{i,p}(\boldsymbol{\theta}, \mathbf{tt}) \right) \tag{25}$$

while the prediction for next time intervals can then be formulated as:

$$X_h^{i+1} = X_h^{i+1,hist} + \sum_{k=h-T}^{h} F_k^h \Delta X_h^i + \left( \sum_{p \in P} \Delta N_p^i P_{h+1}^{i,p}(\boldsymbol{\theta}, \mathbf{tt}) \right) \tag{26}$$

At this stage, the PKF updates the demand for a certain leg $N_p^i$ while keeping the probability $P_h^{i,p}$ constant. Since $P_h^{i,p}$ depends on the travel time, this assumption is reasonable only for small values of $\Delta N_p^i$. This is a strong condition, as it holds only when the contribution of the PKF is negligible. Measurement equation (16) exploits the information from link flows to help reducing this limitation. However, while the transition equation (15) ensures consistency of the demand flows, as discussed in Section 3.2 and showed in system of equations (2), this is not explicitly considered within the measurement equation, meaning that consistency among different demand legs is not guaranteed. For instance, the total demand during the evening commute could be different from the one for the morning commute. In order to avoid this phenomenon, a scale factor $s_p$ can be included in Eq. (22) to introduce artificial consistency in the estimation and prediction phase. For each activity $p$, this weight can be calculated as:

$$s_p = \frac{\sum_{i \in I} N_p^i}{\sum_{i \in I} N_{p-1}^i} \tag{27}$$

where $p$ is the current demand leg, $p$-$1$ is the previous demand leg and $I$ is the set of OD pairs in the network. The main issue with parameter $s_p$ is that, in order to ensure consistency, it will strongly change the value of $\sigma_{PKF}^i$. Specifically, the larger the error within the probability $P_h^{i,p}$, the more $s_p$ will introduce additional errors in the prediction. Moreover, it should be stressed that the PKF propagates errors over time. If the KF achieves a good result, then the PKF will transfer this information to the demand of the next time interval. However, if the KF is providing a biased estimation of the demand, the parameter $s_p$ will propagate this error over time, causing even worse predictions. In this case, the parameter $s_p$ should be adopted only under the condition that the prediction framework adopted to estimate $\Delta X_p^i$ is sufficiently reliable.

*3.6. Input data and comparison to previous works*

As the proposed methodology is based on, but highly differs from conventional state-space formulations, in this section we highlight the contributions, limitations and applications of the proposed PKF, as well as the differences with respect to other existing works.

The proposed PKF is a methodology developed to tackle the *on-line* DODE problem. Specifically, this algorithm exploits prior knowledge about the activity-based structure of the demand to predict its spatial and temporal evolution. Its input data are the (static) purpose-dependent OD flows $\mathbf{N}_p^{hist}$, their spatial structure – i.e. membership map between demand legs and chains of trips, their temporal structure $P_h^{i,p}(\boldsymbol{\theta}, \mathbf{tt})$, and some traffic data $\mathbf{y}_h^{hist}$. In contrast to the proposed approach, the existing algorithms for the on-line DODE usually take as input time-dependent OD flows $\mathbf{x}_h^{hist}$ and traffic data $\mathbf{y}_h^{hist}$ to predict traffic flows. While both approaches provide as output the demand $\mathbf{x}_h^{est}$ and the traffic state $\mathbf{y}_h^{est}$ over the subsequent time intervals, the main difference is that the PKF explicitly targets structural deviations while traditional approaches focus more on local adjustments. Even when structural correlations between variables are explicitly captured (e.g. like in the work of Zhou and Mahmassani (2007)), only temporal correlations are usually considered within on-line DODE models. The proposed PKF accounts for spatial correlations as well.

On the other hand, models that capture spatial correlation between variables have been proposed within the off-line DODE (Cantelmo et al. 2018b; Scheffer, Cantelmo, and Viti 2017). However, these models aim at learning the activity-based structure of the demand from the data in order to generate purpose-dependent aggregated demand patterns. This means that the activity-based structure of the demand, which includes $\mathbf{N}_p^{hist}$ and $\boldsymbol{\theta}$, is an output of the model. Further, these models fail in reproducing accurate traffic predictions, since they do not use real-time traffic data as input. The PKF represents their complement, as it takes $\mathbf{N}_p^{hist}$ and $\boldsymbol{\theta}$ as an input to accurately forecast traffic flows $\mathbf{y}_h^{est}$ over the next time intervals.

## 4. Synthetic Experiments

To show the potential of the novel PKF, we introduce in this section an experiment on a small synthetic network. This experiment uses a quasi-dynamic traffic assignment with limited route choice. These results help in understanding both the potential and limitations of the proposed methodology under controlled settings, hence to show the potential of the proposed PKF in obtaining reliable estimations if compared to the standard KF. The Kalman Filter approach is in fact well known for providing good estimations only for linear systems (Antoniou, Ben-Akiva, and Koutsopoulos 2007). Thus, a linear model is the perfect proof of concept for the proposed methodology. While results with a fully dynamic model are proposed in the next section, the KF model is likely to find a sub-optimal solution, and non-linear versions of the Kalman Filter should be adopted instead. In the next sections, the methodology will be tested with a mesoscopic dynamic traffic model on a large network in order to generalize these findings.

The test network is shown in Fig. 3. The network is composed of five traffic zones, located on Nodes (1,2,3,4,5) and eight bidirectional links. Zones [1-2] are residential, zones [3-4] are working areas and zone [5] is where secondary activities are located. To support the idea that the model can handle multiple and complex mobility patterns, we simulate two different trip chains. Some users travel in the morning from home to work and commute back in the evening. Other users will instead do some leisure activity, passing through zone [5], before going back home.

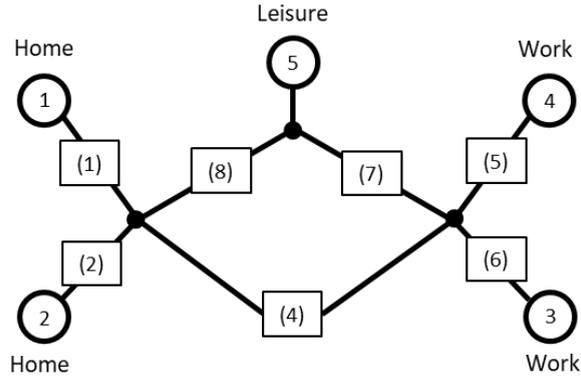

Fig. 3 Experimental Network;

The network has only one detector located on link (4), which captures flows on both directions. We stress that number and location of the detectors influence the performance of the PKF in two ways. First, they influence the measurement equation, as for the normal KF. Secondly, we need to observe an evolution of the demand for the previous demand segment in order to make a prediction. If no evolution of the demand is observed during the morning commute, the PKF will assume that all those demand legs that depend on the morning commute are correct. Beyond this, their number or location does not influence the model performance. Finally, links (8) and (7) have been assigned to be used only by those users going to zone [5], while all remaining demand flows will use link (4) for both morning and evening commute. Link flows are obtained by means of a quasi-dynamic traffic assignment. Specifically, flow propagation is modeled as time dependent, but queues and spillback phenomena are not represented. Travel time is instead generated through simple BPR functions.

In this experiment, the overall demand for the Home-Work morning commuting is set to 26000 people, out of which 6000 users will also perform a Leisure activity before going back home. The time-dependent demand profile is generated through the disutility function proposed by (Arnott, de Palma, and Lindsey 1990). It is also important to stress that the detector located in (4) can properly capture the overall demand for the morning commute but does not provide any information at OD level – i.e. a large number of solutions exists that can reproduce the available traffic data, as it is often the case in practice. Then, we increase the demand by 30%, and use the same departure time choice model to create the historical OD matrix. Eq. (25) is adopted to predict the demand for the next two-time intervals. We test three different approaches: (i) the conventional KF, (ii) the PKF and (iii) the PKF with weight $s_p$ calculated as in Eq (27) in order to ensure conservation of flows (sPKF). Results are reported in Table 1, which shows that all models perform well in reducing the initial error. The KF reduces of 25% the error on the OD flows and 35% the error on the link flows. In terms of traffic measures, similar performances can be observed for the PKF and sPKF. Although the PKF performs slightly better than the others, this improvement is not significant. However, both PKF and sPKF lead to a significant improvement in terms of OD flows with respect to the basic KF. This result was expected, as the KF is relatively reliable when dealing with linear systems, meaning that it manages to find a good approximation for the traffic counts. However, because of the underdeterminedness of the problem, it finds a sub-optimal solution by locally adjusting the OD flows in each time interval. By contrast, the PKF keeps the activity-based structure of the demand, thus reducing significantly the error with respect to the conventional KF in both OD and link flows.

Table 1. Results Synthetic Experiment

| Model Name | RMSE OD Flows | RMSE Link Flows | Improvement OD Flows Prediction | Improvement Link Flows Prediction |
|---|---|---|---|---|
| Seed Matrix | 2676 | 1193 | - | - |
| KF | 1962 | 763 | 26.6 % | 35.9 % |
| PKF+KF | 1722 | 760 | 35.6 % | 36.2 % |
| sPKF+KF | 1701 | 769 | 35.4 % | 35.5 % |

Another important difference lies between the PKF and sPKF. By looking at the aggregate statistics reported in Table 1, the two models seem to provide very similar results. However, the underlying error strongly differs in the two cases because of the assumption on $P_h^{i,p}$ and $s_p$. The estimated profile for the evening commute for the OD pair [4-1] is shown in Fig. 4. Specifically, Figure 4 (b-d) shows the profile estimated through the PKF and the sPKF. Figure 4 (c-e) shows the profile obtained using the KF to further improve this estimation, thus considering both $\sigma_{KF}^i$ and $\sigma_{PKF}^i$.

The reason for choosing OD pair [4-1] is that the evening commute from work to home is where we observed the largest error related to the departure time choice probability. The reason is that the real demand (in black) has a peak when the historical demand is still zero. This is related to the adopted departure time choice model, which assumes that people are spread on a larger time period (Arnott, de Palma, and Lindsey 1990). Because of this, there are large fluctuations that are recovered in a few time intervals thanks to the KF. This behavior can be observed for all models based on the KF (Fig. 4a-c-e). Considering instead the PKF and the sPKF, in Fig. 4 (b-d), we can see the different error related to the assumptions on the parameter $s_p$ and, specifically, on over-imposing a strict conservation of flows.

In this experiment, the overall value of $N_p^i$ is estimated more properly when a strict conservation is imposed in the model, thus when the sPKF is adopted. However, because of the error in the probability $P_h^{i,p}$, the model overestimates the demand in those time intervals in which the demand is already high – i.e. it pushes all the demand where the probability is higher. Clearly, this can lead to substantial errors during the estimation. Even when the model is combined with the normal KF, it takes some time to identify the right demand pattern. The normal PKF is instead underestimating the overall value demand $N_p^i$, but at the same time it provides a more realistic approximation of the demand decreasing the error related to the departure time choice probability function.

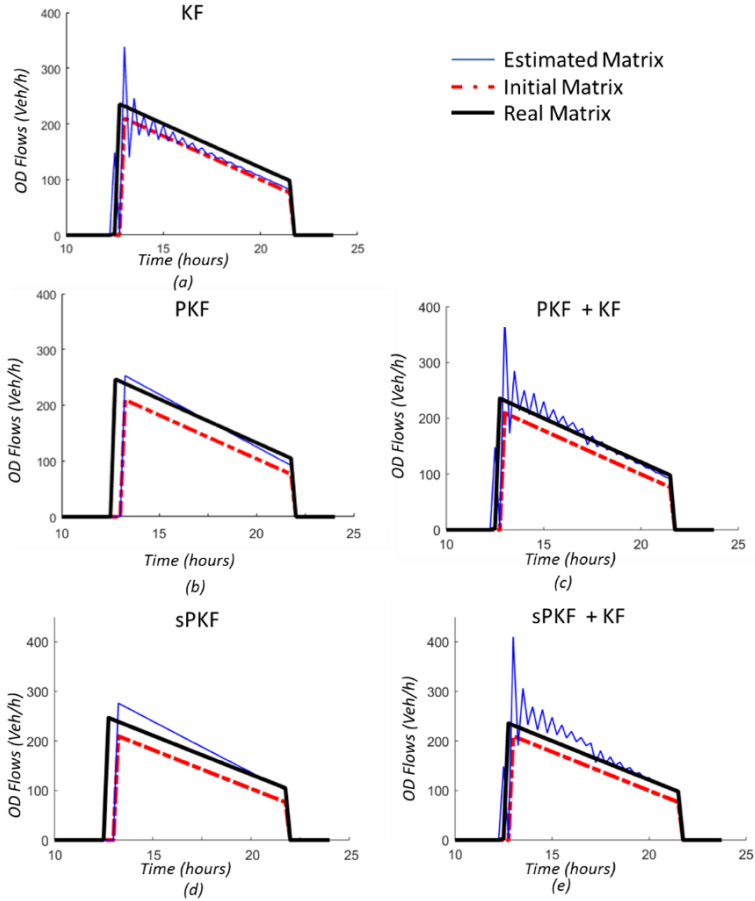

Fig. 4 Estimated Demand, real demand and initial demand for the OD pair (1-4) in the case of (a) KF model, (b) PKF model, (c) PKF and KF,(d) sPKF and (e) sPKF + KF;

## 5. Validation on a large network

### 5.1. Experimental Settings

We now validate the PKF approach to the real large-sized network of Vitoria, Spain. The network is shown in Fig. 5. Vitoria is the capital of the Basque Autonomous Community in northern Spain and represents the typical middle-sized European city in terms of dimension and structure, composed of a city center, a motorway, and suburb areas.

To be able to solve the on-line DODE on this network, we adopted Aimsun (2017) as mesoscopic simulation model, which takes care of computing the assignment matrix through a Dynamic Traffic Assignment process. Differently from the previous synthetic experiment, Aimsun is a commercial software adopted by practitioners all over the world for both planning and real-time traffic management.

The network consists of 2884 nodes, 5799 links, 57 traffic zones and 395 detectors. Simulations are run at mesoscopic resolution, with stochastic route choice scenario (10 replications) and path assignment fixed through dynamic user equilibrium. In this experiment, a 24h demand comprising of 229.646 trips is simulated.

The total demand profile is shown in Fig. 6. Specifically, we simulate two activity patterns, each of them characterized by two legs: the Home-Work-Home activity pattern (in red) and the Home-Leisure-Home activity pattern (in blue). In this case, the departure time probability has been obtained again through the same departure time choice model adopted in the previous synthetic experiment (Arnott, de Palma, and Lindsey 1990).

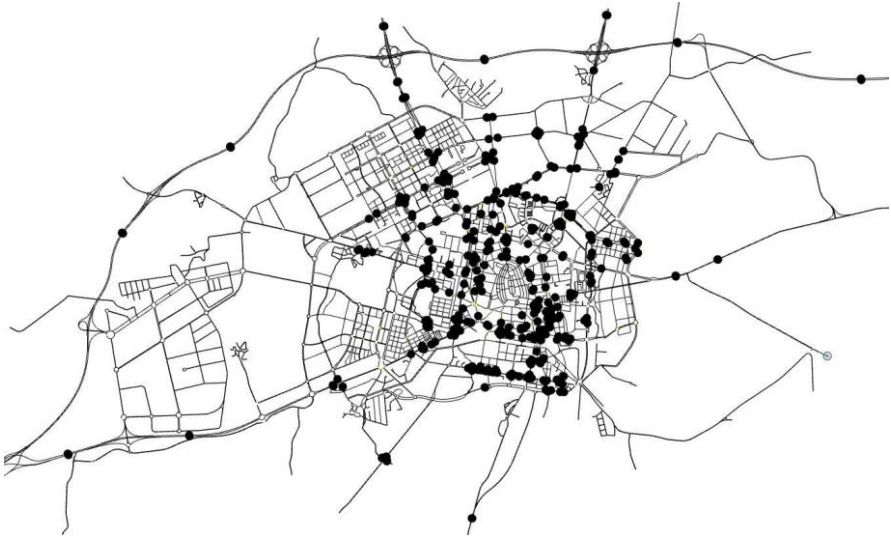

Fig. 5 Network of Vitoria, Spain;

To generate the real matrix, we increased the demand for each trip purpose (i.e. for each leg *p*) by 15%, adding a random noise of 15% with respect to this value for each OD pair. Then, in order to obtain a dynamic matrix, we used again the departure time choice model presented in (Arnott, de Palma, and Lindsey 1990), obtaining a total demand of 275.574 trips. We stress that same parameters for the departure time choice model have been used in order to generate both the "real matrix" and the "historical" – or seed - matrix. These parameters are assumed to be available from the historical database and are adopted to generate the departure time probability $P_h^{i,p}$. However, the resulting value of $P_h^{i,p}$ differs since the travel time on the network changes together with the demand.

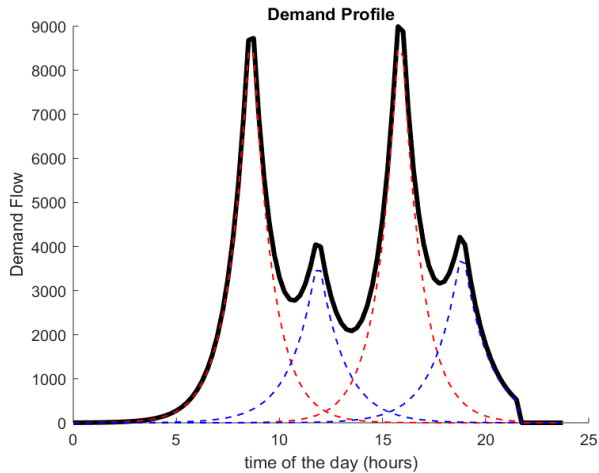

Fig. 6: Demand Profile: Total demand (black); Home-Work-Home commute (red) and Home-Leisure-Home activity pattern (blue);

At this point, we stress that the probability distribution for each activity in the morning and in the evening has a similar shape. However, the probability $P_h^{i,p}$ is an input for the model, and not an output from the estimation. In this sense, the key element is not how similar the probability between morning and evening commute are but how different the adopted value of $P_h^{i,p}$ is with respect to the real one.

These settings have been chosen for three main reasons. First, the network is not congested, as the KF would likely fail in capturing congestion dynamics. Second, the error between the real and historical demand is low. The reason is that while large errors in the demand can be easily captured, in real-time settings, because of e.g. weather conditions and special events, we expect the structural component to change. We can assume this change to be around 15-20%. Finally, the probabilities $P_h^{i,p}$ for different demand legs overlap in most of the time intervals. Again, when the probability during each time interval is fully explained by one single trip purpose, the PKF is capable of transferring the estimation of demand from one time interval to the others. In this instance we would just estimate morning or evening commute as separated instances. However, in reality, demand in each time interval is a convolution of different activity patterns. In Fig. 6, we can see that the demand for leisure activities in the morning overlaps with the demand for both morning commute and evening commute, thus showing how the model performs when the demand is indeed the convolution of different demand segments.

Thus, in this experiment we use KF to estimate the demand for the morning, from 7 AM to 12 AM. Then, we exploit the probability functions $P_h^{i,p}$ to estimate the value of the demand for each activity and we use the PKF to predict the demand in the afternoon. In essence, PKF should help the mobility planner to anticipate traffic problems without waiting for the traffic data to provide this information (e.g. anticipating unusual peaks in the afternoon on the basis of morning observations). Moreover, this approach allows us to see the additional value of the PKF, as the predictions will be based on the same number of simulations and the same traffic data. Thus, we can evaluate results when we only use the KF to estimate local changes in the demand and when we combine this estimation with PKF in order to have a prediction of how demand will evolve during the 24 hours.

*5.2. Experimental results*

Experimental results are summarized in Table 2. We first analyze the standard KF. While the model provides good results in terms of link flows, it fails in reproducing the target matrix. In principle, by considering that we are using the KF for real-time management - this is not a problem, as the estimated matrix provides better predictions (~30% more reliable during the morning commute) with respect to the seed matrix. However, as we pointed out in the previous section, estimated demand during the morning is an input for the PKF in order to update the afternoon demand.

Since the KF runs from 7 am to 12 am, prediction has been done also on the first time intervals of the afternoon. Moreover, link flows depend on the demand generated in the previous time intervals. For this reason, an improvement in the link flows can be observed also in the afternoon (~11%).

Table 2. Results Vitoria Network

| Model Name | RMSE OD Flows | RMSE Link Flows | Improvement OD Flows Prediction | Improvement Link Flows (24h) | Improvement Link Flows (7 AM-12 PM) | Improvement Link Flows (12 PM- 5 PM) |
|---|---|---|---|---|---|---|
| Seed Matrix | 568 | 6910 | - | - | | |
| KF | 700 | 6793 | -9.4 % | 13.18 % | 30.1 % | 11.68 % |
| PKF+KF | 676 | 5072 | -5.7 % | 27.49 % | 43.3 % | 26.06 % |

Concerning the PKF, results confirm the conclusions from the synthetic experiment discussed in section 4. Additionally, Table 2 shows that the error in OD flows for PKF+KF slightly decreases. This is related to two aspects. First, because of the transition equation, the model does not simply calculate the transpose of the demand during the morning. In fact, the probability to propagate demand from one leg to the other is assumed to be uniform. This can be seen as a Markovian process, where for different times of the day we estimate the probability of observing users moving from one demand matrix to another. This means that the demand that arrives to a certain destination in the morning is uniformly distributed among all destinations in the evening. For instance, if zone *k* received 20% more demand, this demand will be distributed among all the ODs using distribution values of the historical demand matrix, regardless where these vehicles are coming from. This is the same concept already shown in system of Equations 3. As a consequence, PKF leads also to better results in terms of link flows. Since some of the demand departs before 12 PM (Fig.6), PKF also improves the prediction in the morning commute.

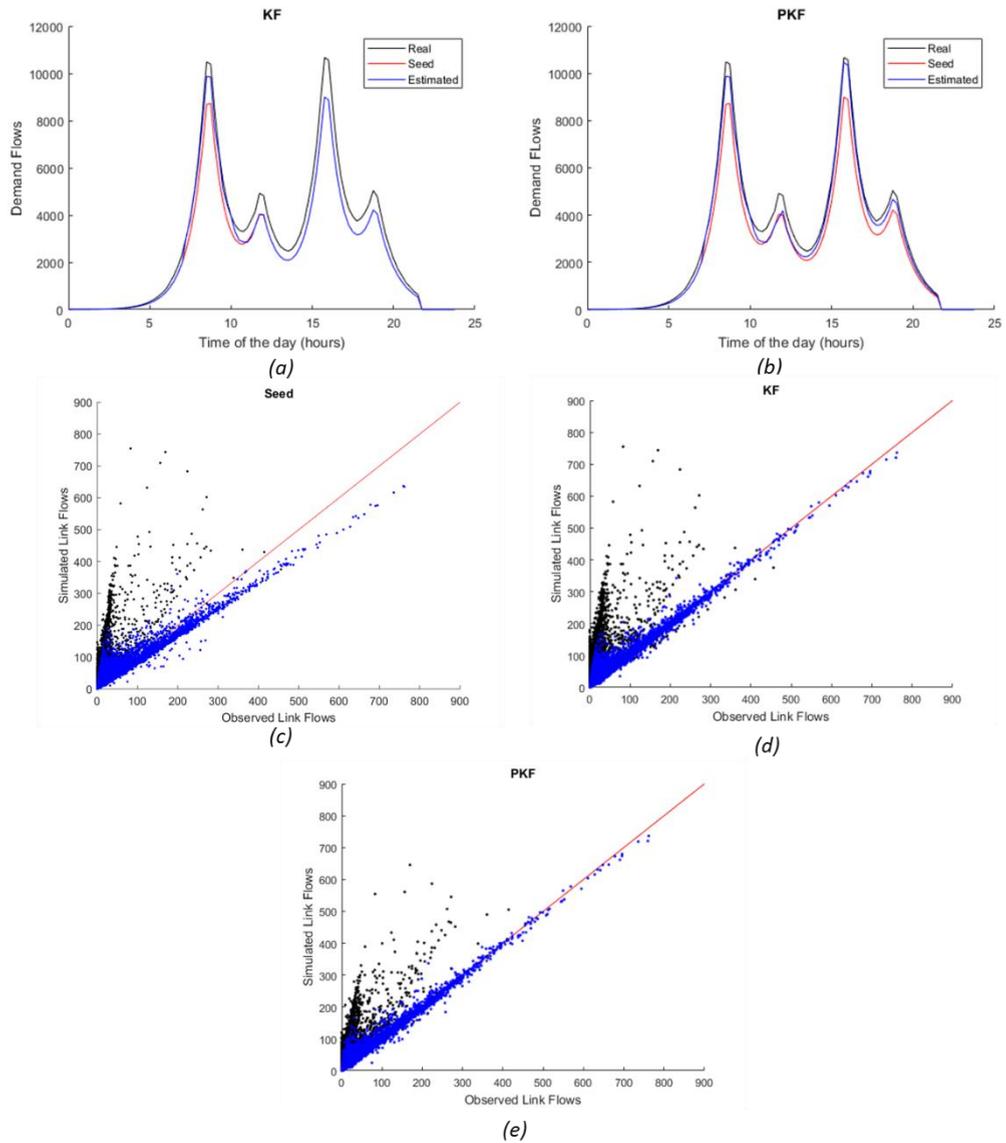

Fig. 7: Scatter plot simulated vs observed traffic counts for the morning commute (blue) and evening commute (black) for the (a) seed matrix, (b) KF , (c) PKF and (d) sPKF;

This suggests that different assumptions on the probability can propagate the error/gain back and forth in time. Second, by not using a strong constraint, but relying on the properties of the Kalman Filter to leverage both the transition equation and the measurement equation, PKF manages to largely increase the model prediction capabilities during the afternoon. This is shown in Figure 7 (a-b), where the total demand estimated with both KF and PKF models is analyzed. First, we can see that the PKF manages to properly identify changes in the demand during the evening commute. Figure 7b shows that the PKF properly captures the peak in the evening commute, as this information has been learned by analyzing the morning rush hour.

This result is more evident when comparing the simulated flows against the real traffic counts (Fig. 7). Specifically, the blue dots represent traffic counts during the morning, while the black dots represent the traffic counts during the afternoon. Although the perturbation is relatively small, Fig. 7c shows that simulated traffic counts for the seed matrix significantly differ from the real ones. The KF model (Fig. 7d) properly solves the issue during the morning commute, as expected, while there is very limited improvement during the afternoon. The PKF (Fig 7e) uses this information to improve the prediction during the afternoon. While the fit for the evening commute is still poor, this result is achieved without need of any further simulation. The prediction is based only on the morning traffic data and the OD demand estimated through the KF. This is also shown in a very intuitive way in Fig. 7 (a-b), which shows the total profile for the estimated demand, according to both KF and PKF. As one can see, the advantage of using the parametric approach is particularly evident for the first hours of the afternoon, when the online model correctly predicts both the mid-day activity and the beginning of the afternoon rush hour.

## 6. Conclusions

This paper introduced a new methodology to consider activity patterns and in particular trip chaining within the online Dynamic OD Estimation process. The methodology, called Parametric Kalman Filter (PKF), combines a departure time choice model with the conventional Kalman Filter (KF) in order to account for activity purpose within the estimation process and find correlation between different OD pairs in time and space.

First, the demand matrix is decomposed in demand legs and demand probabilities. Each demand leg represents the static OD demand for a certain activity purpose. By multiplying the demand of one leg for the probability of departing during a certain time interval, dynamic purpose dependent matrices are generated. The overall demand is then calculated as the convolution of these matrices. This representation of the demand allows calculating transition probabilities from one leg to the other - i.e. the probability of observing the same user travelling during different time intervals.

In order to exploit this information within the online DODE, these transition probabilities were included within a Kalman Filter model. Specifically, the transition equation is formulated in terms of transition probabilities and OD flows for each leg. In this way, the demand observed during previous time intervals can be used to predict trips that will occur later during the day. One of the main issues is related to the error propagation. If a wrong estimation of the demand is available, this error will be transferred to the next time intervals. To avoid this issue, the measurement equation of the Kalman Filter is also modified. Instead of measuring the error for one specific time interval, we calculate the error with respect to the cumulative of the link flows. This equation controls the error propagation by correcting the demand estimated through the transition equation. However, this also leads to have different demand values over different demand legs. In order to control this phenomenon, a parameter was included to explicitly consider the conservation of the demand over time. However, results suggest that the normal PKF, which does not include this constraint, is more reliable in avoiding the error propagation over time while still allowing the model to estimate future traffic patterns based on trip chains dynamics.

Properties of the model were first discussed through a synthetic network and then generalized through an experiment on a real, large-sized network. As the PKF predicts the demand from one leg to the others, it does not require additional computational time with respect to standard models.

The proposed model brings thus the following practical and scientific contributions.
1. The PKF allows modelling trip chain and activity-based demand within the real time OD flows without need to explicitly map activity locations. Only a consistent initial matrix is required;
2. Differently from other models in the literature, which have been proposed for the offline DODE, it does not require a 24h simulation in order to be implemented. If an estimation of the morning rush hour is available, all demand legs depending on this demand can be updated offline and without the need to run additional simulations. Moreover, static assignment matrices can be used in this case, since we need to map the cumulative of the link flows to the state vector, instead of time dependent OD flows;
3. The PKF increases the observability of the demand, since correlation between different OD pairs in time and space is explicitly considered;
4. The PKF helps having results that are more consistent over the 24h periods.

Future research will focus on further testing this methodology under more general conditions. First, the proposed methodology should be tested with non-linear models, such as the Extended Kalman Filter (EKF)(Antoniou, Ben-

Akiva, and Koutsopoulos 2007), as this framework usually provides more reliable estimation with respect to the conventional KF. An additional challenge is to adopt more traffic data, rather than on traffic counts alone. This is a fundamental requirement to test the PKF when severe congestion occurs. An important aspect is also to perform a sensitivity analysis of the model parameters. Specifically, it is important to study how different errors can propagate over time and, even more important, how these errors can be mitigated. In this sense, it is fundamental to investigate the effect of the departure time probability on a day to day basis. Finally, although the proposed formulation is general, the methodology should be tested with more complex activity patterns in order to test the effect of these assumptions on the overall prediction.

## Acknowledgements

We would like to thank Aimsun for its support and for providing the Vitoria network. This work is financially supported by the EU-FEDER project MERLIN and the European Union's Horizon 2020 research and innovation programme under the Marie Skłodowska-Curie grant agreement No 754462..